\def\ra{\rangle}
\def\la{\langle}
\def\be{\begin{equation}}
\def\ee{\end{equation}}
\def\ba{\begin{array}}
\def\ea{\end{array}}

\def\ra{\rangle}
\def\la{\langle}

\def\qed{\leavevmode\unskip\penalty9999 \hbox{}\nobreak\hfill
     \quad\hbox{\leavevmode  \hbox to.77778em{%
               \hfil\vrule   \vbox to.675em%
               {\hrule width.6em\vfil\hrule}\vrule\hfil}}
     \par\vskip3pt}

\documentclass[pra,showpacs,twocolumn,amsmath]{revtex4}
\usepackage{epsfig}
\usepackage{amsmath}

\newtheorem{theorem}{Theorem}
\newtheorem{lemma}{Lemma}

\begin{document}
\title{Inequalities Detecting Quantum Entanglement for $2\otimes d$ Systems}
\author{Ming-Jing Zhao$^{1}$}
\author{Teng Ma$^{2}$}
\author{Shao-Ming Fei$^{1,3}$}
\author{Zhi-Xi Wang$^{1}$}
\affiliation{$^1$School of Mathematical Sciences, Capital Normal
University, Beijing 100048, China\\
$^2$ Department of Physics, Capital Normal University, Beijing
100048, China\\
$^3$Max-Planck-Institute for Mathematics in the Sciences, 04103
Leipzig, Germany}

\begin{abstract}
We present a set of inequalities for detecting quantum entanglement
of $2\otimes d$ quantum states. For $2\otimes 2$
and $2\otimes 3$ systems, the inequalities give rise to sufficient
and necessary separability conditions for both pure and mixed
states. For the case of $d>3$, these inequalities are necessary
conditions for separability, which detect all entangled states that are not positive under partial transposition
and even some entangled states with positive partial transposition. These inequalities are given by
mean values of local observables and present an
experimental way of detecting the quantum entanglement of $2\otimes d$ quantum states
and even multi-qubit pure states.
\end{abstract}

 \maketitle

\section{Introduction}

Entanglement is one of the most fascinating features of quantum
theory and has numerous applications in quantum information
processing \cite{M.A.Nielsen}. Characterization and quantification
of quantum entanglement have become an very important issue. As a
result, various approaches have been proposed and many significant
conclusions have been derived in detecting entanglement \cite{Peres
A., M. Horodecki, M. Lewenstein, M. Horodecki1999, M. Nielsen,T. Hiroshima,P. Horodecki1997,K.Chen}.
One of the most well-known results is the positive partial transpose (PPT) criterion \cite{Peres A., M.
Horodecki}, which says that if a state $\rho$ is separable, then it is
positive under partial transposition. This criterion is both
sufficient and necessary for the separability of qubit-qubit
($2\otimes 2$) and qubit-qutrit ($2\otimes 3$) mixed states. A state
that is not positive under partial transposition is called an NPT
state. It is obvious that PPT criterion can detect all NPT
entangled states but fails in detecting PPT entanglement. The reduction criterion
\cite{M. Horodecki1999} is necessary and sufficient only for $2\otimes 2$ and $2\otimes 3$ states.
Like the majorization criterion \cite{M. Nielsen,T. Hiroshima}, it can neither detect the
PPT entangled states. The range \cite{P. Horodecki1997} and
realignment criteria \cite{K.Chen} are able to detect some PPT entanglement. But generally,
there are yet no general sufficient and necessary separability criteria for higher dimensional states.

Theoretically if one can calculate the degree of entanglement for a
given state, the separability problem can be also solved. For
bipartite systems, there are many well known entanglement measures
such as entanglement of formation
\cite{BDSW,Horo-Bruss-Plenioreviews}, concurrence \cite{con},
negativity \cite{G.Vidal} and relative entropy \cite{V. Vedral}.
However with the increasing dimensions of the systems the
computation of most entanglement measures become formidably
difficult. Therefore many approaches have been used to give an
estimation of the lower bound for entanglement of formation and
concurrence \cite{lowerbound}, which give rise to some necessary
conditions for separability of high dimensional bipartite mixed
states.

For unknown quantum states, the separability can only be determined
by measuring some suitable quantum mechanical observables. The Bell
inequalities \cite{Bell} can be used to detect perfectly the entanglement of pure bipartite
states \cite{CHSH,etc}. Nevertheless these Bell inequalities do not detect
the entanglement of mixed states in general. There are mixed entangled states
which do not violate the Clauser-Horne-Shimony-Holt (CHSH) inequality \cite{werner}. Besides Bell
inequalities, the entanglement witness could also be used for
experimental detection of quantum entanglement
for some special states \cite{M. Horodecki, Philipp05, M. Lewenstein01}.
For two-qubit pure sates, a method to measure the concurrence has been proposed
\cite{F. Mintert2005}, which is further experimentally demonstrated
\cite{S. P. Walborn2006,S. P. Walborn2007}. This protocol needs
a twofold copy of the two-qubit state at every measurement.
A way of measuring concurrence for two-qubit states by using only one copy of the state at
each measurement has been presented in \cite{zmj-concurrence}.
Nevertheless up to now, we have no experimental methods to detect quantum entanglement
sufficiently and necessarily for general mixed states.
Although the PPT criterion is both necessary and sufficient for detecting entanglement of
$2\otimes 2$ and $2\otimes 3$ mixed states, it can not be simply ``translated" into the language
of Bell inequalities.
In \cite{sixiayu}, by using a nice approach and the PPT criterion,
a Bell-type inequality has been proposed for
detecting entanglement of two-qubit mixed states.

In fact the higher dimensional systems offer advantages such as
increased security in a range of quantum information protocols
\cite{hsec}, greater channel capacity for quantum communication \cite{hgc},
novel fundamental tests of quantum
mechanics \cite{hnt}, and more efficient quantum gates \cite{hqg}.
In particular, hybrid qubit-qutrit system has been extensively studied
and already experimentally realized \cite{23e}.
However the approach used in
\cite{sixiayu} can not be simply generalized to the case for $2\otimes d$ systems.

In this paper, we present a set of Bell-type inequalities for
$2\otimes d$ systems, in the sense of \cite{sixiayu} such that the
quantum mechanical observales to be measured are all local ones. We
show that these inequalities can detect all NPT entangled states and
some PPT entangled states. For the separability of $2\otimes 2$ and
$2\otimes 3$ mixed states, these Bell-type inequalities are both
sufficient and necessary. The inequalities can also be used to
detect quantum entanglement experimentally for multiqubit systems.

The paper is organized as follows. In section II, we derive the inequality to detect
entanglement of $2\otimes 3$ system and show that the violation of this inequality implies
quantum entanglement sufficiently and necessarily. Applying our approach to $2\otimes 2$ system, we recover the
main results in \cite{sixiayu}. In section III, we provide inequalities to detect
entanglement for $2\otimes d$ systems and show that these inequalities can detect entanglement of all NPT states
and some PPT states. Conclusions and remarks are given in section IV.

\section{Inequalities for $2\otimes 3$ systems}

First we present a lemma that will be used in proving our theorem for $2\otimes 3$ system.

\begin{lemma}
If the inequality
\begin{eqnarray}\label{lemma1.1}
a_i^2 \geq b_i^2 +c_i^2
\end{eqnarray}
holds for arbitrary real numbers $b_i$ and $c_i$, and nonnegative
$a_i$, $i=1, \cdots, n$, then
\begin{eqnarray*}
(\sum_{i=1}^n p_i a_i)^2 \geq (\sum_{i=1}^n p_i b_i)^2
+(\sum_{i=1}^n p_i c_i)^2
\end{eqnarray*}
for $0 \leq p_i \leq 1$ and $\sum_{i=1}^n p_i=1$.
\end{lemma}

Proof. From Eq. (\ref{lemma1.1}), we have $a_i^2 a_j^2\geq (b_i^2 +
c_i^2 )( b_j^2+ c_j^2 ) \geq (b_i b_j +c_i c_j)^2$. Due to $a_i\geq
0$ for $i=1, \cdots, n$, one gets
\begin{eqnarray*}
(\sum_{i=1}^n p_i a_i)^2 &=& \sum_{i=1}^n p_i^2 a_i^2 + 2
\sum_{i\neq j} p_ip_j a_ia_j \\
&\geq& \sum_{i=1}^n p_i^2 (b_i^2+c_i^2) + 2 \sum_{i\neq j} p_ip_j
(b_ib_j + c_ic_j) \\
&=& (\sum_{i=1}^n p_i b_i)^2 +(\sum_{i=1}^n p_i c_i)^2,
\end{eqnarray*}
which completes the proof of the lemma.\qed

Now let $H_d$ denote a $d$-dimensional vector space with computational
basis $|0\ra=(1,0,...,0)^T$, $|1\ra=(0,1,...,0)^T$, ...,
$|d-1\ra=(0,0,...,1)^T$, where $T$ denotes transpose.
Consider bipartite mixed states in $H_2\otimes H_3$. Let $A_i= U
\sigma_i U^\dagger$, $i=1,2,3$, be a set of quantum mechanical
observables with $U$ any $2\times 2$ unitary matrix, and
$\sigma_1=|0\rangle\langle1| + |1\rangle\langle0|$,
$\sigma_2=i|0\rangle\langle1| -i |1\rangle\langle0|$ and
$\sigma_3=|0\rangle\langle0| - |1\rangle\langle1|$
the Pauli matrices, where $|k\ra\in H_2$, $k=0,1$.
Let $B_j= V \lambda_j V^\dagger$,
$j=1,2,3,4$, be the observables associated with the space $H_3$,
with $V$ any $3\times 3$ unitary matrix, $\lambda_1=|0\rangle\langle0| -
|1\rangle\langle1|$, $\lambda_2=|0\rangle\langle0| -
|2\rangle\langle2|$, $\lambda_3=|0\rangle\langle1| +
|1\rangle\langle0|$ and $\lambda_4=i|0\rangle\langle1| -
i|1\rangle\langle0|$, where $|k\ra\in H_3$, $k=0,1,2$.
According to these observables we can construct inequalities
detecting entanglement perfectly for $2\otimes 3$ system.

\begin{theorem}\label{th2by3}
Any state $\rho$ in $H_2\otimes H_3$ is separable if and only
if the following inequality
\begin{eqnarray}\label{theorem 2by3 *}
&&\langle 2 I_2 \otimes I_3 - I_2 \otimes B_1 +2 I_2 \otimes B_2 +3
A_3 \otimes B_1 \rangle_{\rho} \\[2mm]\nonumber
&&\geq(\langle
3I_2\otimes B_1 +2A_3 \otimes I_3 - A_3 \otimes B_1 + 2 A_3 \otimes
B_2 \rangle^2_{\rho}
\\[2mm]\nonumber
&& + 9 \langle A_1 \otimes B_3 +A_2 \otimes
B_4\rangle^2_{\rho})^{\frac{1}{2}}
\end{eqnarray}
holds for all set of observables $\{A_i\}_{i=1}^3$ and
$\{B_j\}_{j=1}^4$, where $I_d$ denotes the $d\times d$ identity matrix.
\end{theorem}

Proof. Part 1. First we prove that the state is separable if the
inequality (\ref{theorem 2by3 *}) holds. Any pure state
$|\psi\rangle\in H_2\otimes H_3$ has the Schmidt decomposition:
\be\label{psi}
|\psi\rangle = \alpha |00\rangle + \beta|11\rangle,~~~~0 \leq \beta\leq \alpha \leq 1.
\ee
Applying partial transpose with respect
to the first space $H_2$ to $|\psi\rangle\la \psi|$, we get that the corresponding density
matrix $|\psi\rangle\la \psi|$ becomes
$$
|\psi\rangle \langle
\psi|^{T_1} = \alpha^2 |00\rangle\langle 00| +\beta^2
|11\rangle\langle 11| + \alpha\beta(|10\rangle\langle01| +
|01\rangle\langle10|).
$$
By expanding the partial transposed matrix
$|\psi\rangle \langle \psi|^{T_1}$ according to the matrices
$\{\sigma_i\}_{i=1}^3$ and $\{\lambda_j\}_{j=1}^4$ defined above, we get
\be\label{th 2by3 pt}
\ba{l}
|\psi\rangle \langle \psi|^{T_1}\\[2mm]
=\frac{1}{6} (I_2 \otimes I_3 +
(-\frac{1}{2} + \frac{3}{2}\sqrt{1-C^2}) I_2\otimes \lambda_1+
I_2\otimes \lambda_2 \\[2mm]
+ \sqrt{1-C^2} \sigma_3 \otimes
I_3 + (\frac{3}{2}-\frac{1}{2}\sqrt{1-C^2})\sigma_3 \otimes
\lambda_1\\[2mm]
+ \sqrt{1-C^2} \sigma_3 \otimes \lambda_2) + \frac{1}{4}C
(\sigma_1\otimes \lambda_3 + \sigma_2\otimes\lambda_4),
\ea
\ee
where $C=2\alpha \beta$ is just the concurrence of the pure state
$|\psi\rangle$, defined by $C(|\psi \rangle)= \sqrt{2(1-Tr \rho_1^2)}$.
$\rho_1$ is the reduced density matrix $\rho_1=Tr_2(|\psi
\rangle \langle \psi|)$, where $Tr_2$ stands for the partial trace with respect to
the second space.

Let $U$ be an arbitrary $2\times 2$ unitary matrix
and $V$ an arbitrary $3\times 3$ unitary matrix. Then $|\Psi\ra \equiv U^* \otimes V
|\psi\rangle $ represents an arbitrary pure state in $H_2\otimes
H_3$. Note that a bipartite state $\rho\in H_2\otimes H_3$ is
separable if and only if $\rho^{T_1}$ is positive, that is,
$\la\Psi|\rho^{T_1}|\Psi\ra\geq 0$  for all $|\Psi\ra \in
H_2\otimes H_3$. Therefore
$$\ba{rcl}
0&\leq&\langle \psi |U^T \otimes
V^\dagger \rho^{T_1} U^* \otimes V |\psi\rangle\\[2mm]
&=&Tr(\rho^{T_1} U^*
\otimes V |\psi\rangle \langle \psi| U^T \otimes V^\dagger)\\[2mm]
&=& Tr(\rho U \otimes V (|\psi\rangle \langle \psi|)^{T_1} U^\dagger \otimes
V^\dagger) \\[2mm]
&\equiv& \langle U \otimes V (|\psi\rangle \langle
\psi|)^{T_1} U^\dagger \otimes V^\dagger \rangle_{\rho}
\ea
$$
for all $U$, $V$, $\alpha$ and $\beta$, where $Tr(A^{T_1} B)= Tr
(AB^{T_1})$ has been taken into account and $Tr$ stands for trace.
Hence we have
\begin{eqnarray}\label{2by3 12negativity}
&&12\la\Psi|\rho^{T_1}|\Psi\ra\\\nonumber
&&=12\langle U \otimes V (|\psi\rangle \langle \psi|)^{T_1} U^\dagger
\otimes V^\dagger \rangle_{\rho} \\\nonumber &&= \langle 2 I_2
\otimes I_3 + (-1 + 3\sqrt{1-C^2}) I_2\otimes B_1 + 2I_2\otimes B_2
\\\nonumber &&+ 2\sqrt{1-C^2} A_3 \otimes I_3 + (3-\sqrt{1-C^2})A_3 \otimes
B_1
\\\nonumber&&+ 2\sqrt{1-C^2} A_3 \otimes B_2\rangle_{\rho}
+ 3C \langle A_1\otimes B_3 + A_2\otimes
B_4\rangle_{\rho}\\\nonumber &&\geq \langle 2 I_2 \otimes I_3  -
I\otimes B_1 + 2 I_2\otimes B_2 + 3A_3 \otimes
B_1\rangle_{\rho}\\\nonumber
 &&-| \sqrt{1-C^2} \langle 3I_2\otimes B_1 +2 A_3
\otimes I_3 - A_3 \otimes B_1 \\\nonumber &&+ 2A_3 \otimes B_2
\rangle_{\rho} + 3C \langle A_1\otimes B_3 + A_2\otimes B_4
\rangle_{\rho}|\\\nonumber &&\geq \langle 2 I_2 \otimes I_3  -
I_2\otimes B_1 + 2 I_2\otimes B_2 + 3A_3 \otimes
B_1\rangle_{\rho}\\\nonumber &&-\{ \langle 3I_2\otimes B_1 +2 A_3
\otimes I_3 - A_3 \otimes B_1 + 2A_3 \otimes B_2 \rangle^2_{\rho}
\\\nonumber && +9 \langle A_1\otimes B_3 + A_2\otimes B_4
\rangle^2_{\rho}\}^{\frac{1}{2}},
\end{eqnarray}
where we have used Eq. (\ref{th 2by3 pt}) and employed the definition
of $\{A_i\}$ and $\{B_j\}$ for the second equality.
The first inequality is due to $-|x|\leq x$ and the second one
is from the Cauchy inequality. Therefore if the inequality
(\ref{theorem 2by3 *}) holds, the right hand side of
inequality (\ref{2by3 12negativity}) is nonnegative. Therefore
$\la\Psi|\rho^{T_1}|\Psi\ra\geq 0$ for all $|\Psi\ra\in H_2\otimes
H_3$, and the state is separable according to the PPT criterion.

Part 2. We prove now that if the state is separable, the inequality
(\ref{theorem 2by3 *}) holds. First we show that inequality
(\ref{theorem 2by3 *}) holds for all pure separable states, which is
equivalent to prove that for arbitrary pure separable state $\rho$,
the following inequality holds:
\begin{eqnarray}\label{theorem 2by3 **}
&&\langle 2 I_2 \otimes I_3 - I_2 \otimes \lambda_1 +2 I_2 \otimes
\lambda_2 +3 \sigma_3 \otimes \lambda_1 \rangle^2_{\rho} \\\nonumber
&&\geq \langle 3I_2\otimes \lambda_1 +2\sigma_3 \otimes I_3 -
\sigma_3 \otimes \lambda_1 + 2 \sigma_3 \otimes \lambda_2
\rangle^2_{\rho}
\\\nonumber
&& + 9 \langle \sigma_1 \otimes \lambda_3 +\sigma_2 \otimes
\lambda_4\rangle^2_{\rho}.
\end{eqnarray}

Note that any pure separable state can be written as $|\xi\rangle= (\gamma_1
|0\rangle + \gamma_2|1\rangle)\otimes ( \phi_0 |0\rangle + \phi_1 |1\rangle +\phi_2
|2\rangle)$ with $|\gamma_1|^2+|\gamma_2|^2=1$ and $|\phi_0|^2 +|\phi_1|^2 +|\phi_2|^2=1$.
Inserting this separable pure state $|\xi\rangle \langle \xi|$ into Eq. (\ref{theorem 2by3 **}),
one gets that the square root of the left hand side of (\ref{theorem 2by3 **}) becomes
\be \label{2by3 pure ai>0}
\ba{l} \langle 2 I_2 \otimes I_3 - I_2 \otimes \lambda_1 +2 I_2
\otimes \lambda_2 +3 \sigma_3 \otimes \lambda_1
\rangle_{|\xi\rangle \langle \xi|}\\[1mm]
=6(|\phi_0\gamma_1|^2+|\phi_1\gamma_2|^2)\geq 0. \ea \ee While the right hand
side of the inequality (\ref{theorem 2by3 **}) becomes
\be\label{2by3 a
not 0 right} \ba{l} \langle 3I_2\otimes \lambda_1 +2\sigma_3 \otimes
I_3 - \sigma_3 \otimes \lambda_1 + 2 \sigma_3 \otimes \lambda_2
\rangle^2_{|\xi\rangle \langle \xi|}
\\[1mm]
+ 9 \langle \sigma_1 \otimes \lambda_3 +\sigma_2 \otimes
\lambda_4\rangle^2_{|\xi\rangle \langle \xi|}\\[1mm]
=(6|\gamma_1 \phi_0|^2-6|\gamma_2\phi_1|^2)^2
+144(Re(\gamma_1\gamma_2^*)Re(\phi_1^* \phi_0))^2.
\ea
\ee
The difference between the left and right hand side of (\ref{theorem 2by3 **}) is given by
\be\ba{l}
\langle 2 I_2 \otimes I_3 - I_2 \otimes \lambda_1 +2 I_2 \otimes
\lambda_2 +3 \sigma_3 \otimes \lambda_1
\rangle^2_{|\xi\rangle \langle \xi|}
\\[2mm]
-\langle 3I_2\otimes \lambda_1 +2\sigma_3 \otimes I_3
- \sigma_3 \otimes \lambda_1 + 2 \sigma_3 \otimes \lambda_2
\rangle^2_{|\xi\rangle \langle \xi|}
\\[2mm]
- 9 \langle \sigma_1 \otimes \lambda_3 +\sigma_2 \otimes
\lambda_4\rangle^2_{|\xi\rangle \langle \xi|}\\[2mm]
=144 |\gamma_1\gamma_2 \phi_0\phi_1|^2-144(Re(\gamma_1\gamma_2^*)Re(\phi_1^* \phi_0))^2\geq0.
\ea \ee
Therefore the inequality (\ref{theorem 2by3 **}) holds for any pure separable states.

We now prove that the inequality (\ref{theorem 2by3 *})
also holds for general separable mixed states,
$$
\rho=\sum_i p_i |\psi_i\rangle \langle \psi_i|,~~~0\leq p_i\leq 1,~~~\sum_i p_i =1,
$$
where
$|\psi_i\rangle$ are all pure separable states.
Set
$$
\ba{l}
a_i=\langle 2 I_2 \otimes I_3 - I_2 \otimes B_1 +2 I_2 \otimes B_2
+3 A_3 \otimes B_1 \rangle_{|\psi_i\rangle\langle \psi_i|},\\[2mm]
b_i=\langle
3I_2\otimes B_1 +2A_3 \otimes I_3 - A_3 \otimes B_1 + 2 A_3 \otimes
B_2 \rangle_{|\psi_i\rangle\langle \psi_i|},\\[2mm]
c_i= 3\langle A_1 \otimes B_3
+A_2 \otimes B_4\rangle_{|\psi_i\rangle\langle \psi_i|}.
\ea
$$
We have
$$
\ba{l}
(\sum_i p_i a_i)^2\\[1mm]
=\langle 2 I_2 \otimes I_3 - I_2 \otimes B_1 +2 I_2 \otimes B_2
+3 A_3 \otimes B_1 \rangle^2_{\rho},\\[2mm]
(\sum_i p_ib_i)^2\\[1mm]
=\langle 3 I_2 \otimes B_1 + 2 A_3 \otimes I_3 - A_3 \otimes B_1
+2 A_3 \otimes B_2 \rangle^2_{\rho},\\[2mm]
(\sum_i p_i c_i)^2
=9\langle A_1 \otimes B_3 + A_2 \otimes B_4\rangle^2_{\rho}.
\ea
$$

Since inequality
(\ref{theorem 2by3 **}) holds for all pure separable states,
$a_i^2 \geq b_i^2 +c_i^2$. Furthermore, from the
inequality (\ref{2by3 pure ai>0}) one gets $a_i\geq 0$.
From the lemma one gets $(\sum_i p_i a_i)^2 \geq (\sum_i p_i
b_i)^2 +(\sum_i p_i c_i)^2$, which verifies that any mixed
separable state $\rho$ obeys the inequality (\ref{theorem 2by3 *}).
\qed

We have shown that any state $\rho$ in $H_2\otimes H_3$ is separable if and only
if the inequality (\ref{theorem 2by3 *}) is satisfied.
The inequality (\ref{theorem 2by3 *}) gives a necessary and sufficient separability criterion
for general qubit-qutrit states. The separability of the state can be determined by experimental measurements
on the local observables. For instance, we
consider the mixed state
$$
\rho=p|\psi^+\rangle
\langle \psi^+| +\frac{1-p}{6}I_6,
$$
where $|\psi^+\rangle=\frac{1}{\sqrt{2}}(|00\rangle +|11\rangle)$.
Let us take $U=I_2$ and $V=
|0\rangle \langle 1| + |1\rangle \langle 0| + |2\rangle \langle 2|$.
Let $F^{(3)}_{\{U\},\{V\}}(\rho)$ denote the value of violation of the
inequality (\ref{theorem 2by3 *}),
\begin{eqnarray}\label{2by3 violation}
&&F^{(3)}_{\{U\},\{V\}}(\rho)\\\nonumber&&\equiv (\langle
3I_2\otimes B_1 +2A_3 \otimes I_3 - A_3 \otimes B_1 + 2 A_3 \otimes
B_2 \rangle^2_{\rho}
\\\nonumber &&+ 9 \langle A_1 \otimes B_3 +A_2 \otimes
B_4\rangle^2_{\rho})^{\frac{1}{2}}\\\nonumber &&-\langle 2 I_2
\otimes I_3 - I_2 \otimes B_1 +2 I_2 \otimes B_2 +3 A_3 \otimes B_1
\rangle_{\rho}.
\end{eqnarray}
By straightforward calculation we have
$F_{U,V}^{(3)}(\rho)=8p-2>0$ for
$p>\frac{1}{4}$. As this state is entangled if and only if
$p>\frac{1}{4}$, our inequality (\ref{theorem 2by3 *}) detects all
the entanglement of the state.

We consider now the maximal violation of the inequality (\ref{theorem 2by3 *}).
Let $F^{(3)}(\rho)=\max_{\{U\}, \{V\}} \{F^{(3)}_{\{U\},
\{V\}}(\rho), 0\}$ denote the maximal violation value with respect
to a given state $\rho$, under all $\{U\}$ and $\{V\}$. Obviously, $F^{(3)}(\rho)=0$ if
$\rho$ is separable.  For an entangled state $\rho$, $F^{(3)}(\rho)\geq
-12\lambda_{\min}$, where $\lambda_{\min}$ is the minimal
eigenvalue of the partial transposed density matrix of $\rho$,
$\lambda_{\min}= \min_{U,V, \alpha, \beta} \langle U\otimes V
(|\psi\rangle \langle\psi|)^{T_1} U^\dagger \otimes
V^\dagger\rangle_{\rho}$, where $|\psi\ra$ is given by Eq. (\ref{psi}).
As an example, let us simply take the
observables $\{A_i\}_{i=1}^3$ to be $\{\sigma_1, \sigma_2,
\sigma_3\}$ and $\{B_j\}_{j=1}^4$ to be $\{ \lambda_1, \lambda_2,
\lambda_3, \lambda_4 \}$, i.e. $U=I_2$ and $V=I_3$ in the theorem
\ref{th2by3}. The violation corresponding to the pure state $\alpha_0 |01\rangle
+\beta_0|10\rangle$ is
$F_{I_2,I_3}^{(3)}(\rho)=12\alpha_0 \beta_0$. For the maximally entangled state
$\frac{1}{\sqrt{2}} (|01\rangle +|10\rangle)$, the
corresponding maximal violation value is 6.

For given $U$ and $V$, inequality (\ref{theorem 2by3 *})
also gives rise to a kind of entanglement witness $W_{U,V}$:
\begin{eqnarray*}
W_{U,V}&=&\langle 2 I_2 \otimes I_3 - I_2 \otimes B_1 +2 I_2 \otimes B_2 +3
A_3 \otimes B_1 \rangle_{\rho} \\\nonumber
&-&(\langle
3I_2\otimes B_1 +2A_3 \otimes I_3 - A_3 \otimes B_1 + 2 A_3 \otimes
B_2 \rangle^2_{\rho}\\\nonumber
&+& 9 \langle A_1 \otimes B_3 +A_2 \otimes
B_4\rangle^2_{\rho})^{\frac{1}{2}}.
\end{eqnarray*}
For all separable states $\sigma$, $Tr(W_{U,V}\sigma)\geq 0$. If $Tr(W_{U,V}\rho)<0$ then $\rho$ is entangled.
Every entanglement witness $W_{U,V}$ detects a certain set of entangled states.
Witnesses $\{W_{U,V}\}$ under all $U$ and $V$ together detect all the entangled
states, since all entangled states violate the inequality (\ref{theorem 2by3 *}).

Here as in \cite{sixiayu}, the observables in theorem \ref{th2by3} are not
independent. The three observables $\{A_i\}_{i=1}^3$
for the first subsystem and the four observables $\{B_i\}_{j=1}^4$ for the
second subsystem fulfill the relations $A_1A_2=-iA_3$ and $B_3B_4=-iB_1$
respectively.

Based on the PPT criterion, we have derived the inequality which is both
sufficient and necessary for separability of $H_2\otimes H_3$ system.
Our approach can be also applied to other systems such as two-qubit ones, although
the approach used in \cite{sixiayu} can not be simply applied to $H_2\otimes H_3$ system.
In term of our approach it is easily to get the following result for two-qubit system:
{\it Any two-qubit state $\rho$ is separable if and only if
\begin{eqnarray}\label{cor2by2 *}
&&\langle  I_2 \otimes I_2 + A_3 \otimes B_3 \rangle_{\rho}
\\\nonumber &&\geq (\langle I_2\otimes B_3 +A_3 \otimes I_2 \rangle^2_{\rho}
+ \langle A_1 \otimes B_1 +A_2 \otimes
B_2\rangle^2_{\rho})^{\frac{1}{2}}
\end{eqnarray}
for all set of observables $\{A_i\}_{i=1}^3$ and $\{B_j\}_{j=1}^3$,
where $A_i= U \sigma_i U^\dagger$ and $B_j= V \sigma_j V^\dagger$,
$i,j=1,2,3$, $U$ and $V $ are $2\times 2$ unitary matrices.}
The observables here have the same
orientation $\mu =-iA_1 A_2 A_3=-iB_1 B_2 B_3=1$.
If one replaces $\sigma_3$ with $-\sigma_3$, the above
inequality still holds. But the
orientation becomes $\mu =-iA_1 A_2 A_3=-iB_1 B_2 B_3=-1$. Namely
the inequality (\ref{cor2by2 *}) is true for all set of observables
with the same orientation, which recover the results in  \cite{sixiayu}.
Moreover, one can also obtain that,
for a given entangled state the maximal violation of the
inequality (\ref{cor2by2 *}) is $-4\lambda_{\min}$. The
possible maximal violation among all states is 3, which is attainable by the
maximally entangled states \cite{sixiayu}.

\section{Inequalities for $2\otimes d$ systems}

For higher dimensional bipartite systems, the PPT criterion is only
necessary for separability. In the following we study the Bell-type
inequalities for $H_2\otimes
H_d$ systems. The quantum states in $H_2\otimes
H_d$ also play important roles in quantum information
processing \cite{P. Horodecki, B. Kraus, B. Bylicka}. The
separability for $H_2\otimes H_d$ systems could also shed light on the
separability of multiqubits systems.

\begin{theorem}
(i) Any separable state $\rho\in H_2\otimes H_d$ obeys the following
inequality:
\begin{eqnarray}\label{cor2byd *}
&&\langle 2 I_2 \otimes I_d + (2-d)I_2 \otimes B_1 +2 I_2 \otimes
B_2 +\cdots \\\nonumber && + 2 I_2 \otimes B_{d-1} + d A_3 \otimes
B_1 \rangle_{\rho}
\\\nonumber &&\geq (\langle dI_2\otimes B_1 +2A_3 \otimes I_d +(2-d) A_3
\otimes B_1 + 2 A_3 \otimes B_2 \\\nonumber &&+\cdots + 2 A_3
\otimes B_{d-1} \rangle^2_{\rho}
+ d^2 \langle A_1 \otimes B_d +A_2 \otimes
B_{d+1}\rangle^2_{\rho})^{\frac{1}{2}},
\end{eqnarray}
where the obserbables $\{A_i\}_{i=1}^3$ are defined as the ones in
theorem \ref{th2by3}. $B_j=V\lambda_j V^\dagger$, $j=1,\cdots, d+1$, with $V$ any
$d\times d$ unitary matrix, $\lambda_1=|0\rangle\langle0|
   -|1\rangle\langle1|$, $\lambda_2=|0\rangle\langle0|
   -|2\rangle\langle2|$, $\cdots$, $\lambda_{d-1}=|0\rangle\langle0|
   -|d-1\rangle\langle d-1|$, $\lambda_{d}=|0\rangle\langle 1|
   +|1\rangle\langle 0|$ and $\lambda_{d+1}=i|0\rangle\langle1|
   -i|1\rangle\langle0|$, $|j\ra\in H_d$, $j=0,...,d-1$.

(ii) All NPT states in $H_2\otimes H_d$ violate the above inequality.
\end{theorem}

The proof of (i) is similar to the part 2 in the proof of theorem
\ref{th2by3} for necessity of separability. The statement (ii) can
be proved analogous to the part 1 in the proof of theorem
\ref{th2by3}. However as the PPT criterion is no longer both
sufficient and necessary for separability of $2\otimes d$ systems,
one has only that all NPT entangled states violate the inequality.

For the cases $d=2$ and $d=3$, the inequality (\ref{cor2byd
*}) reduces to the inequality (\ref{cor2by2 *}) and
(\ref{theorem 2by3 *}) respectively.

Let $F^{(d)}(\rho)$ denote the maximal violation value of the
inequality (\ref{cor2byd *}) for a given state $\rho$: $F^{(d)}(\rho)=\max_{\{U\}, \{V\}} \{F^{(d)}_{\{U\},
\{V\}}(\rho), 0\}$, where
\begin{eqnarray}
&&F^{(d)}_{\{U\},
\{V\}}(\rho)\\\nonumber
&&=(\langle dI_2\otimes B_1 +2A_3 \otimes I_d +(2-d) A_3
\otimes B_1 + 2 A_3 \otimes B_2 \\\nonumber &&+\cdots + 2 A_3
\otimes B_{d-1} \rangle^2_{\rho}
+ d^2 \langle A_1 \otimes B_d +A_2 \otimes
B_{d+1}\rangle^2_{\rho})^{\frac{1}{2}}\\\nonumber
&&-\langle 2 I_2 \otimes I_d + (2-d)I_2 \otimes B_1 +2 I_2 \otimes
B_2 +\cdots \\\nonumber && + 2 I_2 \otimes B_{d-1} + d A_3 \otimes
B_1 \rangle_{\rho}.
\end{eqnarray}
Analogously,
we have that $F^{(d)}(\rho)$ is invariant under local unitary
transformations, $F^{(d)}(\rho)=F^{(d)}(U\otimes V\rho U^\dagger
\otimes V^\dagger)$ and $F^{(d)}(\rho)=0$ if $\rho$ is
separable. For any entangled state $\rho$, we have $F^{(d)}(\rho)
\geq -4d\lambda_{\min}$, where $\lambda_{\min}$ is the minimal
eigenvalue of the partial transposed density matrix of $\rho$. Any
violation of the inequality (\ref{cor2byd *}) implies entanglement.
Since all entangled pure states are NPT, Eq. (\ref{cor2byd *}) can
detect all pure entangled states. Moreover, as all mixed states with
rank less than or equal to $d$ are entangled if and only if they
are NPT \cite{P. Horodecki}, inequality (\ref{cor2byd *}) can also detect
the entanglement of all such states.

An interesting thing is that although inequality (\ref{cor2byd *})
is obtained based on PPT criterion which is no longer
sufficient for separability of $2\otimes d$ systems for $d>3$, it can still detect
the quantum entanglement of some PPT
entangled states. Namely, besides all NPT states, some PPT entangled
states would also violate the inequality. This can be seen from the
proof of the first part of the theorem \ref{th2by3}. Any PPT state $\rho$ satisfies
$\la\Psi|\rho^{T_1}|\Psi\ra\geq 0$ for all pure state $|\Psi\ra$.
From Eq. (\ref{2by3 12negativity}) one can similarly obtain that, for
$2\otimes d$ systems, it is possible that the inequality
(\ref{cor2byd *}) is violated while $\la\Psi|\rho^{T_1}|\Psi\ra\geq
0$ is still satisfied. As an example we consider the family of PPT
entangled states in $2\otimes 4$ systems, introduced in \cite{P.
Horodecki1997}:
\be\label{eg} \sigma_b=
\frac{7b}{7b+1}\sigma_{insep} + \frac{1}{7b+1} |\phi_b\rangle
\langle \phi_b|,
\ee
where
$$\ba{l}
\sigma_{insep}=
\frac{2}{7}(|\psi_1\rangle \langle \psi_1| + |\psi_2\rangle \langle
\psi_2|+ |\psi_3\rangle \langle \psi_3|)+\frac{1}{7}|14\rangle
\langle 14|,\\[2mm]
|\phi_b\rangle= |2\rangle\otimes
(\sqrt{\frac{1+b}{2}}|1\rangle + \sqrt{\frac{1-b}{2}}|3\rangle ),\\[2mm]
|\psi_1\rangle= \frac{1}{\sqrt{2}}(|11\rangle +
|22\rangle),\\[2mm]
|\psi_2\rangle= \frac{1}{\sqrt{2}}(|12\rangle +
|23\rangle),\\[2mm]
|\psi_3\rangle= \frac{1}{\sqrt{2}}(|13\rangle +
|24\rangle),
\ea
$$
where $0\leq b\leq 1$. The state $\sigma_b$ is entangled if and only if $0<b<1$ \cite{P. Horodecki1997}.

In fact, we can simply choose $U=|0\rangle \langle 1| - |1\rangle \langle 0|$ and $V=I_4$.
Then $F_{\{I_2\},\{I_4\}}^{(4)}(\sigma_b)=-8b - 4(1 + b) + \sqrt{4096 b^2 + (-8 b + 4 (1 + b))^2}$
and $F_{\{I_2\},\{I_4\}}^{(4)}(\sigma_b)>0 $ when $\frac{1}{31}<b<1$. Therefore, the
inequality can detect almost all the entanglement in $\sigma_b$ (see FIG. 1).
In deed our inequality has advantages in detecting entanglement of this PPT entangled state,
since the PPT, CCNR, reduction and majorization criteria can all not detect the
entanglement of $\sigma_b$.

\begin{center}
\begin{figure}[!h]\label{fig}
\resizebox{4cm}{!}{\includegraphics{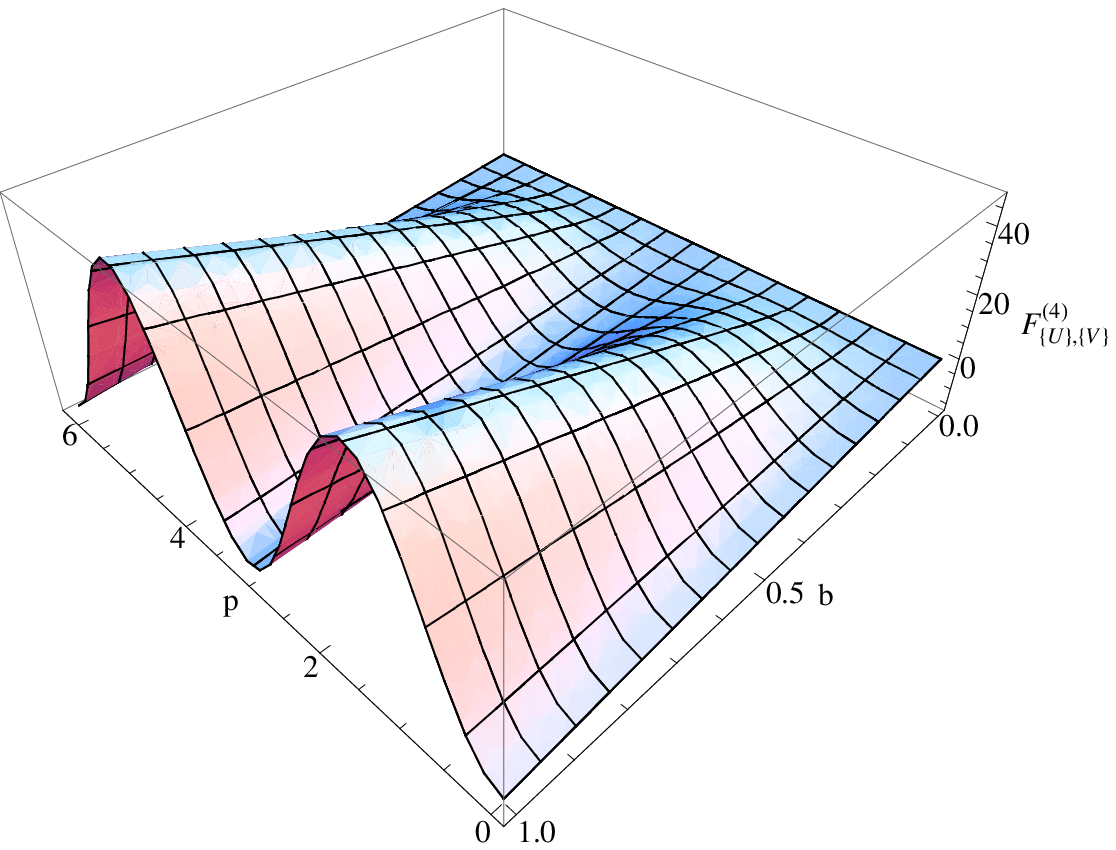}}
\resizebox{3cm}{!}{\includegraphics{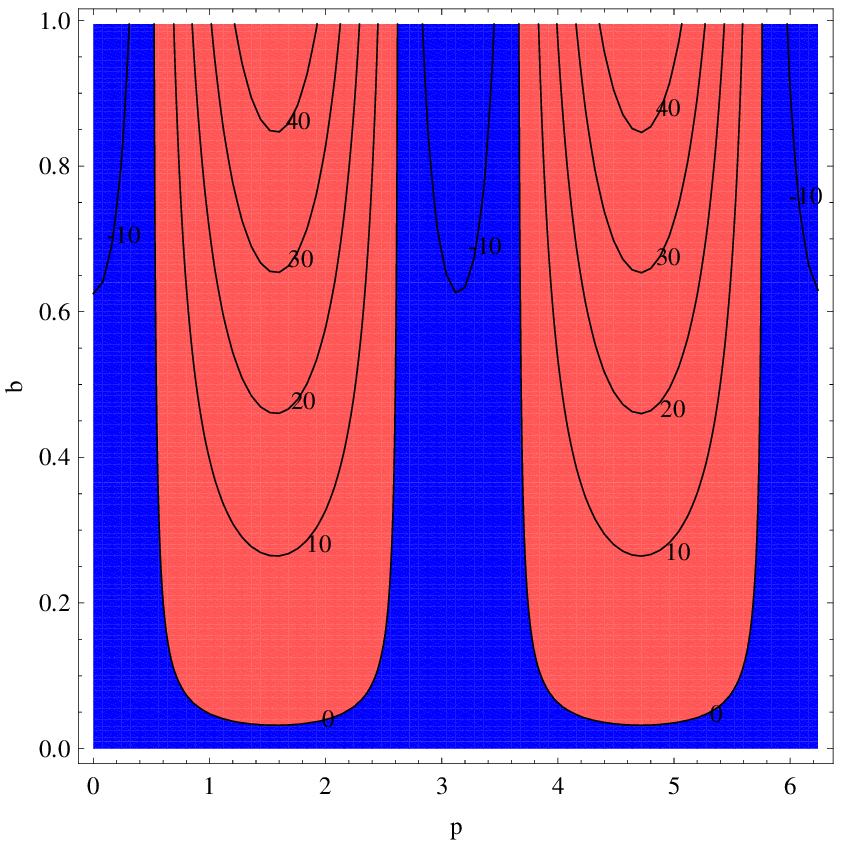}}\caption{$U=\cos p
(|0\rangle \langle 0|+|1\rangle \langle 1|)+\sin p (|0\rangle \langle 1| - |1\rangle \langle 0|)$, $V=I_4$.
Left figure: $F_{\{U\},\{V\}}^{(4)}(\sigma_b)$ with respect to $p$ and $b$. Right figure:
contour plot of the left figure. The dark region: $F_{\{U\},\{V\}}^{(4)}(\sigma_b)<0$;
the gray region: $F_{\{U\},\{V\}}^{(4)}(\sigma_b)>0$.}
\end{figure}
\end{center}

The inequality (\ref{cor2byd *}) can also detect
entanglement for $n$-qubit pure states. Suppose $|\psi\rangle_{A_1
\cdots A_n}$ is an arbitrary $n$-qubit pure state. If we treat the
$n$-qubit state $|\psi\rangle_{A_1 \cdots A_n}$ as a bipartite one
with the $i$-th qubit as one subsystem and the rest qubits as
another subsystem, then it is a $2\otimes 2^{n-1}$ bipartite pure
state. $|\psi\rangle_{A_1 \cdots A_n}$ is separable under this
partition if and only if it fulfills the inequality (\ref{cor2byd *}).

\medskip

\section{Conclusions}

In terms of a new approach we have derived a series of Bell-type
inequalities detecting quantum entanglement for $2\otimes d$ systems. These
inequalities work for both pure and mixed states. All the separable
states obey these inequalities and all NPT entangled states violate them.
They are both sufficient and necessary for
separability of $2\otimes 2$ and $2\otimes 3$ systems. They give rise to an experimental way
to detect the entanglement, as only the mean values of local observables are involved.
These inequalities are a kind of experimental
realization of PPT criterion. But they are more powerful than the PPT criterion, as they can also detect
entanglement of some PPT
entangled states. Our inequalities are complementary to some known separability criteria
for PPT entanglement. In addition, our inequalities also provide an
experimental way of detecting quantum entanglement for multiqubit pure states.

\noindent{\bf Acknowledgments}\, M.J. Zhao thanks S.X. Yu for
valuable discussions. This work is supported by the NSFC 10875081,
NSFC 10871227, KZ200810028013, PHR201007107 and NSF of Beijing
1092008.


\begin{thebibliography}{18}

\bibitem{M.A.Nielsen} M. A. Nielsen, and I. L. Chuang, {\it Quantum Computation and Quantum
Information} (Cambridge University Press, Cambridge, 2000).

\bibitem{Peres A.} A. Peres, Phys. Rev. Lett.  {\bf 77},  1413(1996).

\bibitem{M. Horodecki} M. Horodecki, P. Horodecki, and R. Horodecki, Phys. Lett. A  {\bf 223}, 1(1996).

\bibitem{M. Lewenstein} M. Lewenstein, and A. Sanpera, Phys. Rev. Lett.  {\bf 80},  2261(1998).

\bibitem{M. Horodecki1999} M. Horodecki, and P. Horodecki, Phys. Rev. A \textbf{59}, 4206(1999).

\bibitem{M. Nielsen} M. A. Nielsen, and J. Kempe, Phys. Rev. Lett.  {\bf 86}, 5184(2001).

\bibitem{T. Hiroshima} T. Hiroshima, Phys. Rev. Lett.  {\bf 91}, 057902(2003).

\bibitem{P. Horodecki1997} P. Horodecki, Phys. Lett. A {\bf 232}, 333(1997).

\bibitem{K.Chen}
K. Chen and L. A. Wu. Quantum Inf. Comput. \textbf{3}, 193(2003);\\
O. Rudolph, Quantum Inf. Proc. \textbf{4}, 219(2005).


\bibitem{BDSW} C. H. Bennett, D. P. DiVincenzo, J. A. Smolin, and W. K.
Wootters, Phys. Rev. A \textbf{54}, 3824(1996).

\bibitem{Horo-Bruss-Plenioreviews} M. Horodecki, Quantum Inf. Comp. \textbf{1}, 3(2001);\\
M. B. Plenio, and S. Virmani, Quantum Inf. Comp. \textbf{7},
1(2007).

\bibitem{con}
A. Uhlmann, Phys. Rev. A \textbf{62}, 032307(2000);\\
P. Rungta, V. Bu\v{z}ek, C. M. Caves, M. Hillery, and G. J. Milburn,
Phys. Rev. A \textbf{64}, 042315(2001);\\S. Albeverio, and S. M.
Fei, J. Opt. B: Quantum Semiclass. Opt. \textbf{3}, 223(2001).

\bibitem{G.Vidal} G. Vidal, and R. F. Werner, Phys. Rev. A \textbf{65}, 032314(2002).

\bibitem{V. Vedral} V. Vedral, and M. B. Plenio, Phys Rev A \textbf{57}, 1619(1998).

\bibitem{lowerbound}
F. Mintert, M. Ku\'s, and A. Buchleitner, Phys. Rev. Lett. \textbf{92}, 167902(2004);\\
K. Chen, S. Albeverio, and S. M. Fei,
Phys. Rev. Lett. \textbf{95}, 210501(2005);\\ K. Chen, S. Albeverio, and S. M. Fei, Phys. Rev. Lett. \textbf{95}, 040504(2005);\\
H. P. Breuer,  J. Phys. A: Math. Gen. \textbf{39}, 11847(2006);\\
J. I. de Vicente, Phys. Rev. A \textbf{75}, 052320(2007);\\
X. H. Gao, S. M. Fei, and K. Wu, Phys. Rev. A \textbf{74}, 050303(2006);\\
E. Gerjuoy, Phys. Rev.
A \textbf{67}, 052308(2003);\\
Y. C. Ou, H. Fan, and S. M. Fei, Phys. Rev. A \textbf{78},
012311(2008).

\bibitem{Bell} J. S. Bell, Physics (Long Island, N. Y.) {\bf 1}, 195(1964).

\bibitem{CHSH} J. Clauser, M. Horne, A. Shimony, and R. Holt, Phys. Rev. Lett. {\bf 23}, 880(1969);\\
N. Gisin, Phys. Lett. A {\bf 154}, 201(1991).

\bibitem{etc}
D. Collins, N. Gisin, N. Linden, S. Massar, and S. Popescu, Phys. Rev. Lett. \textbf{88}, 040404 (2002);\\
M. Li, and S. M. Fei, Phys. Rev. Lett. \textbf{104}, 240502 (2010).

\bibitem{werner}R. F. Werner, Phys. Rev. A \textbf{40}, 4277 (1989).

\bibitem{Philipp05} P. Hyllus, O. G$\mathrm{\ddot{u}}$hne, D. Bru$\mathrm{{\ss}}$, and
M. Lewenstein, Phys. Rev. A  {\bf 72}, 012321(2005).

\bibitem{M. Lewenstein01} M. Lewenstein, B. Kraus, P. Horodecki, and J. I. Cirac, Phys. Rev. A  {\bf 63},
044304(2001).

\bibitem{F. Mintert2005} F. Mintert, M. Ku$\rm{\acute{s}}$, and A. Buchleitner,
Phys. Rev. Lett. \textbf{95}, 260502(2005).

\bibitem{S. P. Walborn2006} S. P. Walborn, P. H. Souto Ribeiro, L. Davidovich, F. Mintert, and A. Buchleitner, Nature.
\textbf{440}, 20(2006).

\bibitem{S. P. Walborn2007} S. P. Walborn, P. H. Souto Ribeiro, and L. Davidovich, Phys Rev A \textbf{75}, 032338(2007).

\bibitem{zmj-concurrence}
S. M. Fei, M. J. Zhao, K. Chen and Z. X. Wang, Phys. Rev. A {\bf 80}, 032320(2009).

\bibitem{sixiayu} S. Yu, J. W. Pan, Z. B. Chen, and Y. D. Zhang,
Phys. Rev. Lett. {\bf 91}, 217903(2003).

\bibitem{hsec} N. K. Langford, R. B. Dalton, M. D. Harvey, J. L. O'Brien, G. J. Pryde, A. Gilchrist, S. D. Bartlett, and A. G. White, Phys. Rev. Lett. {\bf 93}, 053601(2004);\\
G. Molina-Terriza, A. Vaziri, J. \v{R}eh\'{a}\v{c}ek, Z. Hradil, and A. Zeilinger, Phys. Rev. Lett. {\bf 92}, 167903(2004);\\
S. Gr\"oblacher, T. Jennewein, A. Vaziri, G. Weihs, and A. Zeilinger, New J. Phys. {\bf 8}, 75(2006);\\
D. Bruss and C. Macchiavello, Phys. Rev. Lett. {\bf 88}, 127901(2002);\\
N. J. Cerf, M. Bourennane, A. Karlsson, and N. Gisin, Phys. Rev. Lett. {\bf 88}, 127902(2002).

\bibitem{hgc} M. Fujiwara, M. Takeoka, J. Mizuno, and M. Sasaki, Phys. Rev. Lett. {\bf 90}, 167906(2003).

\bibitem{hnt} D. Collins, N. Gisin, N. Linden, S. Massar, and S. Popescu, Phys. Rev. Lett. {\bf 88}, 040404(2002).

\bibitem{hqg} T. C. Ralph, K. Resch, and A. Gilchrist, Phys. Rev. A {\bf 75}, 022313(2007).

\bibitem{23e} B. P. Lanyon, T. J. Weinhold, N. K. Langford, J. L. O'Brien, K. J. Resch, A. Gilchrist, and A. G. White, Phys. Rev. Lett. {\bf 100}, 060504(2008);\\
K. Ann, and G. Jaeger, Phys. Lett. A {\bf 372}, 579(2008).

\bibitem{P. Horodecki} P. Horodecki, M. Lewenstein, G. Vidal, and I. Cirac, Phys. Rev. A \textbf{62}, 032310(2000).

\bibitem{B. Kraus} B. Kraus, J. I. Cirac, S. Karnas, and M. Lewenstein, Phys. Rev. A \textbf{61}, 062302(2000).

\bibitem{B. Bylicka} B. Bylicka, and D. Chru\'{s}ci\'{n}ski, Phys. Rev. A \textbf{81}, 062102(2010).



\end{thebibliography}
\end{document}